\documentclass[aps,prl,twocolumn,showpacs,reprint,superscriptaddress,longbibliography]{revtex4-1}

\usepackage{amsmath, amsfonts, amssymb, bm}
\usepackage{mathtools}
\usepackage[inline]{enumitem}
\usepackage{graphicx}
\usepackage{xcolor}
\usepackage[breaklinks,hypertexnames=false]{hyperref}
\hypersetup{colorlinks,linkcolor={magenta},citecolor={blue},urlcolor={blue}}

\usepackage[capitalize]{cleveref}

\usepackage{glossaries}
\glsdisablehyper     

\newacronym{mbs}{MBS}{Majorana bound state}
\newacronym{qsh}{QSHI}{quantum spin Hall insulator}
\newacronym{zbp}{ZBCP}{zero-bias conductance peak}

\begin{document}

\title{Identifying Majorana bound states at quantum spin Hall edges using a metallic probe}

\newcommand{\tianjin}{Center for Joint Quantum Studies and Department of Physics,
	Tianjin University, Tianjin 300072, China}
\newcommand{\uam}{Department of Theoretical Condensed Matter Physics, Condensed Matter Physics Center (IFIMAC) and Instituto Nicol\'as Cabrera, Universidad Aut\'onoma de Madrid, 28049 Madrid, Spain}
\newcommand{\nagoya}{Department of Applied Physics,
	Nagoya University, Nagoya 464-8603, Japan}

\author{Bo Lu}
\affiliation{\tianjin}

\author{Guanxin Cheng}
\affiliation{\tianjin}

\author{Pablo Burset}
\affiliation{\uam}

\author{Yukio Tanaka}
\affiliation{\nagoya}

\date{\today}
\begin{abstract}
	We study the conductance afforded by a normal-metal probe which is directly contacting the helical edge modes of a quantum spin Hall insulator (QSHI). We show a $2e^2/h$ conductance peak at zero temperature in QSHI-based superconductor--ferromagnet hybrids due to the formation of a single Majorana bound state (MBS). In a corresponding Josephson junction hosting a pair of MBSs, a $4e^2/h$ conductance peak is found at zero temperature. The conductance quantization is robust to changes of the relevant system parameters and, remarkably, remains unaltered with increasing the distance between probe and MBSs. In the low temperature limit, the conductance peak is robust as the probe is placed within the localization length of MBSs. Our findings can therefore provide an effective way to detect the existence of MBSs in QSHI systems. 
\end{abstract}

\maketitle

\section{I. INTRODUCTION}
A \gls{qsh} is a two dimensional (2D) topological insulator featuring topologically protected one dimensional (1D) edge states~\cite{Kane051,Kane052,Bernevig06,Liu08,Wu06,Xu06,Hsu_2021}. These edge states are termed \textit{helical} since they circulate in reversed directions with opposite spin orientations. 
\Glspl{qsh} have been realized using HgTe~\cite{Konig07,KonigJ07,Roth09,Brune12} and InAs/GaSb~\cite{Knez11} quantum wells, among others~\cite{Reis17,Kammhuber17,Wu18}, and are attracting significant attention as a platform for the creation and manipulation of \glspl{mbs}; a very promising building-block for fault-tolerant quantum computations~\cite{Kitaev01,Kitaev03,Nayak08,Hasan10,XLQi11,Alicea11,Leijnse12,Ando13,Tkachov13}. 
Signatures of \glspl{mbs} include a fractional, or $4\pi$-periodic, Josephson effect in \gls{qsh}-mediated junctions~\cite{Kwon04,Fu09}, which has been recently observed~\cite{Bocquillon16,Bocquillon17,Deacon17}. Similar signatures of \glspl{mbs} have been predicted in ferromagnetic--superconductor (FS) junctions proximity-induced at the helical edge of a \gls{qsh}, hosting a single \gls{mbs}~\cite{FuL09,Fu09}, or a pair of them for SFS junctions~\cite{Kitaev01,Kwon04,Fu09}. 

Another basic transport signature of helical edge states is a
quantized tunneling conductance in units of $e^{2}/h$, with $e$ being the
electron charge and $h$ the Planck constant. Since the helical edge states of a \gls{qsh} comprise \textit{one half} of a spin degenerate 1D electron gas, each edge can provide a quantized ballistic conductance of $e^{2}/h$ over several micrometers~\cite{Konig07,Brune10,Du15}. 
When coupled to a superconductor, the conductance per edge is doubled ($2e^{2}/h$)~\cite{Knez12} as a result of the perfect Andreev reflection of the helical states~\cite{Adroguer10,Sun11,Narayan12}. 
Consequently, a \gls{mbs} in a \gls{qsh} is predicted to feature a quantized gate-voltage-averaged conductance peak~\cite{Shuo13}. Indeed, experimental evidence of \glspl{mbs} in other topological systems has focused on measuring a robust quantization of \glspl{zbp} ~\cite{Mourik12,Deng12,Lee12,Das12,Deng16,Jelena16,Albrecht16,Nichele17,Gul18,TanakaA04,TanakaA05,TanakaJ12,Cayao_2021,Chiu2021}. 

\begin{figure}
\includegraphics[width=0.45\textwidth]{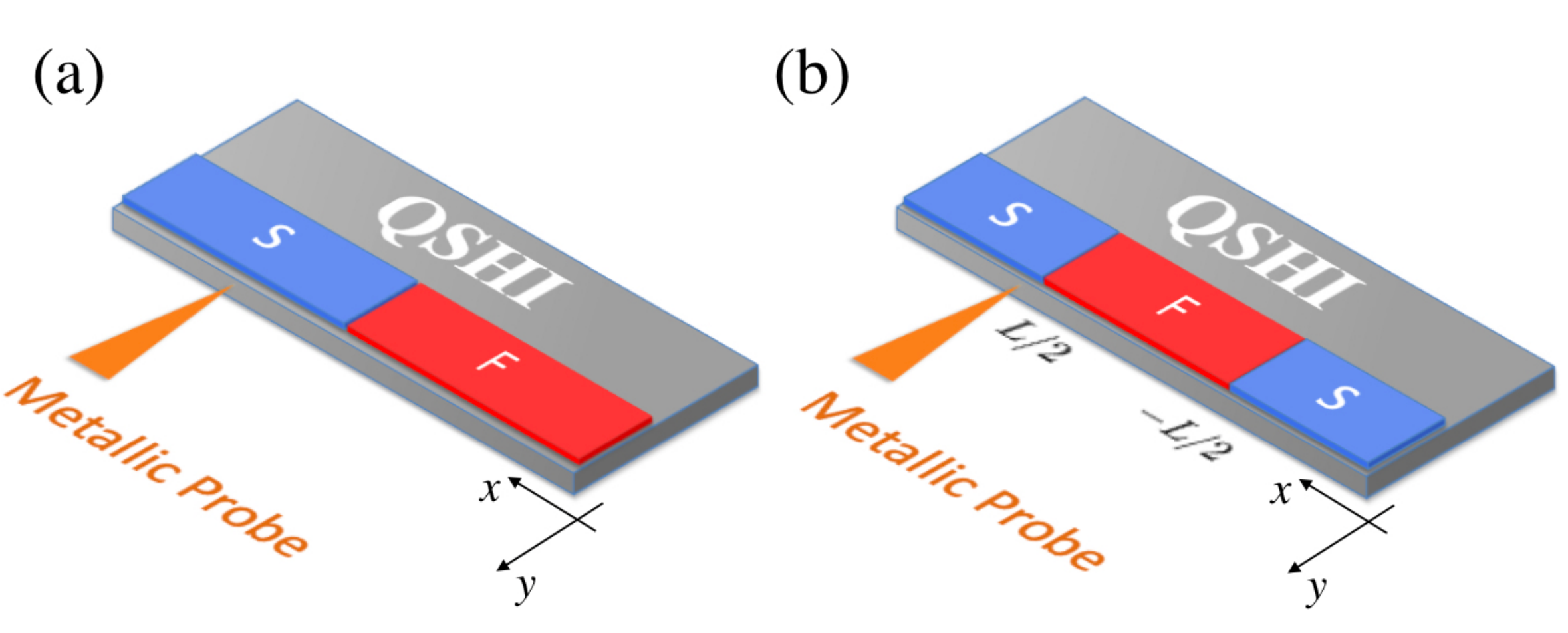}
\caption{Schematics of the three-terminal setup. FS (a) and SFS (b) junctions based on a single \gls{qsh} edge, with a biased one-dimensional normal-metal probe terminal.  }
\label{fig:01}
\end{figure}

In this article, we propose to detect \glspl{mbs} in the helical edge state via a normal probe terminal, such as an STM tip, coupled to the edge states (\cref{fig:01}). We only consider one edge, assuming negligible overlap between different edge states due to a large width of the \gls{qsh}. 
A similar three-terminal setup with a metallic tip has been proposed to detect the helical nature of the edge states or to design novel applications~\cite{Shivamoggi10,Benjamin10,Das11,Sothmann17,Bours18,Bours19,Blasi20,Soori20}. 
Here, we focus on the detection of \glspl{mbs} and consider both superconducting and magnetic regions, which open gaps on the helical edge channel forming the FS and SFS junctions featured in, respectively, \cref{fig:01}(a) and \cref{fig:01}(b). 
In the beam-splitter configuration, a bias voltage $V$ at the probe terminal injects a current $I$ into the helical edge state, propagating along it until is collected by distant grounded contacts. When the FS junction hosts a single \gls{mbs} (or a pair for the SFS junction), the conductance $dI/dV$ at the probe terminal reaches a quantized plateau of $2e^{2}/h$ ($4e^{2}/h$) at zero bias. 
The conductance quantization is robust under the change of all relevant system parameters. Strikingly, the quantization is also independent of the position of the probe terminal along the edge, which is in contrast to the decaying local density of states of \glspl{mbs}. The ZBCP at finite temperature is suppressed due to thermal broadening but remains observable in the low temperature limit as the probe is placed within the localization length of MBSs. Such a zero-bias coherent transport effect thus provides an experimental signature of \glspl{mbs} in a QSHI.

\section{II. MODEL AND FORMULAS}
\emph{Model.---}
Our setup consists of a semi-infinite normal probe on the $y$ axis and a %
\gls{qsh} edge lying along the $x$ direction (\cref{fig:01}). We set the
origin at the FS boundary, or at the middle of the SFS, and place the
contact point between the \gls{qsh} edge and the probe terminal at $\left(
x_{c},0\right)$. The Hamiltonian is $\hat{H}=\frac{1}{2}\Psi ^{\dag }H\Psi $%
, with $H=H_{N}+H_{T}+H_{j}$, $\Psi =(\psi _{\uparrow },\psi _{\downarrow},
\psi _{\uparrow }^{\dag },\psi _{\downarrow }^{\dag })^{T}$, and $\psi
_{\uparrow(\downarrow)}$ the field operators for right (left) movers. $H_{N}$
describes the normal probe terminal with a standard parabolic dispersion
\begin{equation}
	H_{N}=\left( -\frac{\hbar ^{2}\partial _{y}^{2}}{2m_{N}}-\mu _{N}\right)
	\hat{\tau}_{z}\Theta \left( y\right) +U\hat{\tau}_{z}\delta \left(
	y-0^{+}\right) ,  \label{eq1}
\end{equation}
where $m_{N}$, $\mu _{N}$, and $\hat{\tau}_{i=x,y,z}$ are the effective
mass, chemical potential of normal probe, and Pauli matrices in the Nambu
space. $\Theta \left( y\right)$ and $\delta \left( y\right) $ are the
Heaviside and delta functions, respectively, and $U$ is the barrier
parameter between the probe and the \gls{qsh}. The bare helical edge states
are described by the Hamiltonian
\begin{equation}
	H_{T}=-i\hbar v_{t}\hat{\sigma}_{z}\partial _{x}-\mu _{T}\hat{\tau}_{z},
	\label{eq2}
\end{equation}%
where $v_{t}$ is the Fermi velocity, $\mu _{N}$ is the chemical potential
measured from the Dirac point of the helical states, and $\hat{\sigma}%
_{i=x,y,z}$ are Pauli matrices in the spin space. Finally, $H_{j}$ describes
the proximity-induced terms in the helical state with $j=1$ ($j=2$)
corresponding to the FS (SFS) junction. Specifically,
\begin{align}
	H_{1}={}&M_{x}\hat{\sigma}_{x}\hat{\tau}_{z}\Theta \left( -x\right) -\Delta
	\hat{\sigma}_{y}\hat{\tau}_{y}\Theta \left( x\right) ,  \label{eq3} \\
	H_{2} ={}&M_{x}\hat{\sigma}_{x}\hat{\tau}_{z}\Theta \left( \frac{L}{2}
	-\left\vert x\right\vert \right) -\Delta \hat{\sigma}_{y}\hat{\tau}
	_{y}e^{i\chi _{L}}\Theta \left( -x-\frac{L}{2}\right)  \notag  \label{eq4} \\
	&-\Delta \hat{\sigma}_{y}\hat{\tau}_{y}e^{i\chi _{R}}\Theta \left( x-\frac{L
	}{2}\right) .
\end{align}%
Here, $M_{x}$ is an exchange field from the coupling to a ferromagnetic
insulator and $\Delta $ is the proximity-induced pair potential from a
conventional $s$-wave superconductor. For the SFS junction, $L$ is the width
of the $F$ region and $\chi_{L,R}$ are the macroscopic superconducting
phases, with $\chi =\chi _{L}-\chi _{R}$ being their difference.

The scattering wave function for the normal probe is
\begin{align}
	\Psi _{N\sigma } ={}&\Phi _{\sigma }+b_{\uparrow \sigma }\hat{B}_{\uparrow
	}e^{iqy}+b_{\downarrow \sigma }\hat{B}_{\downarrow }e^{iqy}  \notag \\
	&+a_{\uparrow \sigma }\hat{A}_{\uparrow }e^{-iqy}+a_{\downarrow \sigma }
	\hat{A}_{\downarrow }e^{-iqy},
\end{align}
for an incident electron with spin $\sigma=\uparrow(\downarrow)$ and wave
function $\Phi _{\uparrow}=(1,0,0,0)^{T}e^{-iqy}$ [$\Phi _{\downarrow
}=(0,1,0,0)^{T}e^{-iqy}$]. Under the wide band approximation ($\mu_N\gg E$),
the wave vector is $q=\sqrt{2m_{N}\mu _{N}}/\hbar $ and the spinors are $%
\hat{B}_{\uparrow }=(1,0,0,0)^{T}$, $\hat{B}_{\downarrow }=(0,1,0,0)^{T}$, $%
\hat{A}_{\uparrow }=(0,0,1,0)^{T}$, and $\hat{A}_{\downarrow }=(0,0,0,1)^{T}$%
. The normal (Andreev) reflection amplitudes are $b_{\sigma ^{\prime }\sigma
}$ ($a_{\sigma ^{\prime }\sigma }$) for an incoming electron of spin $\sigma
$ scattered as an electron (hole) of spin $\sigma^{\prime }$.

At the edge of \gls{qsh}~\cite%
{Crepin14,Bo15,Crepin15,Burset15,Cayao17,Keidel18}, the solutions on the F
region are $\hat{F}_{1}=N_{1}^{-1}(\hbar v_{t}\kappa _{e}+\mu
_{T},M_{x},0,0)^{T}e^{i\kappa _{e}x}$, $\hat{F}_{2}=N_{2}^{-1}(M_{x},\hbar
v_{t}\kappa _{e}+\mu _{T},0,0)^{T}e^{-i\kappa _{e}x}$, $\hat{F}%
_{3}=N_{3}^{-1}(0,0,\hbar v_{t}\kappa _{h}+\mu _{T},M_{x})^{T}e^{-i\kappa
	_{h}x}$, and $\hat{F}_{4}=N_{4}^{-1}(0,0,M_{x},\hbar v_{t}\kappa _{h}+\mu
_{T})^{T}e^{i\kappa _{h}x}$, with wave vectors $\kappa _{e\left( h\right) }=%
\sqrt{\mu _{T}^{2}-M_{x}^{2}}/\hbar v_{t}$ for $M_{x}<\mu _{T}$, and $\kappa
_{e\left( h\right) }=\pm i\sqrt{M_{x}^{2}-\mu _{T}^{2}}/\hbar v_{t}$ for $%
M_{x}>\mu _{T}$. $N_{i=1\sim 4}$ are normalization factors. On the S region
we find $\hat{S}_{1}=\left( u,0,0,v\right) ^{T}e^{ik^{+}x}$, $\hat{S}%
_{2}=\left( 0,-u,v,0\right) ^{T}e^{-ik^{+}x}$, $\hat{S}_{3}=\left(
0,-v,u,0\right) ^{T}e^{-ik^{-}x}$, and $\hat{S}_{4}=\left( v,0,0,u\right)
^{T}e^{ik^{-}x}$, with $u\left( v\right) =\sqrt{(E\pm \sqrt{E^{2}-\Delta
		^{2} })/2E}$, wave vector $k^{\pm }=k_{T}\pm \sqrt{E^{2}-\Delta ^{2}}/(\hbar
v_{t})$ and $k_{T}=\mu _{T}/(\hbar v_{t})$. An incident spin-$\sigma $
electron from the normal probe thus has a transmitted wave function $\Psi
_{T\sigma }\left( x\right) $ on the \gls{qsh} side, which is a suitable
superposition of spinors $\hat{F}_{i=1\sim 4}$ and $\hat{S}_{i=1\sim 4}$.

To determine the scattering amplitudes, we match $\Psi _{N\sigma }$ and $%
\Psi _{T\sigma }$ at the contact using the boundary conditions~\cite%
{Modak12,Soori13,Soori20}
\begin{align}
	\Psi _{N\sigma }\left( y=0^{+}\right) ={}& c\Psi _{T\sigma }\left(
	x_{c}^{-}\right) +c\Psi _{T\sigma }\left( x_{c}^{+}\right) , \\
	\hat{K}\Psi _{N\sigma }\left( y=0^{+}\right) ={}& iv_{t}\left[ \Psi
	_{T\sigma }\left( x_{c}^{-}\right) -\Psi _{T\sigma }\left( x_{c}^{+}\right) %
	\right] ,
\end{align}%
with $\hat{K}=\hbar c\hat{\sigma}_{z}\hat{\tau}_{z}\left( m_{N}^{-1}\partial
_{y}-2U/\hbar ^{2}\right) $. The parameter $Z=m_{N}U/(\hbar ^{2}q)$
describes the barrier strength between the probe and the helical edge, while
the real and dimensionless number $c$ represents the different microscopic
details between them, such as the hopping integrals in the underlying
lattice model~\cite{Modak12,Soori13,Soori20}.

When a bias voltage $V$ is applied to the normal probe, the conductance $G$
at zero temperature is calculated using the formula~\cite{BTK}
\begin{equation}
	G=\frac{e^{2}}{h}\sum\nolimits_{\sigma \sigma ^{\prime }}\left( \delta
	_{\sigma \sigma ^{\prime }}+\left\vert a_{\sigma \sigma ^{\prime
	}}\right\vert ^{2}-\left\vert b_{\sigma \sigma ^{\prime }}\right\vert
	^{2}\right) .  \label{eq13}
\end{equation}
Here, $a_{\sigma \sigma ^{\prime }}$ and $b_{\sigma \sigma ^{\prime }}$ are
reflection amplitudes at $E=eV$.

For the numerical calculations we choose $v_{N}\approx 1.57\times 10^{6}$
m/s and $\mu _{N}\approx 7.0$ eV, corresponding to copper; $v_{t}\approx
5.5\times 10^{5}$ m/s and $\mu _{T}\approx 0.01$ eV for HgCdTe quantum
wells; and $\Delta \approx 0.1$ meV, which corresponds to a
proximity-induced gap from a niobium superconductor. For simplicity, we only
consider the short junction for the SFS geometry, with $k_{T}L\ll \mu
_{T}/\Delta $, and choose $k_{T}L=2$ so that the width of F is $L\approx 70$ nm.
We note that our main conclusions are not material dependent.

\begin{figure}
	\includegraphics[width=0.47\textwidth]{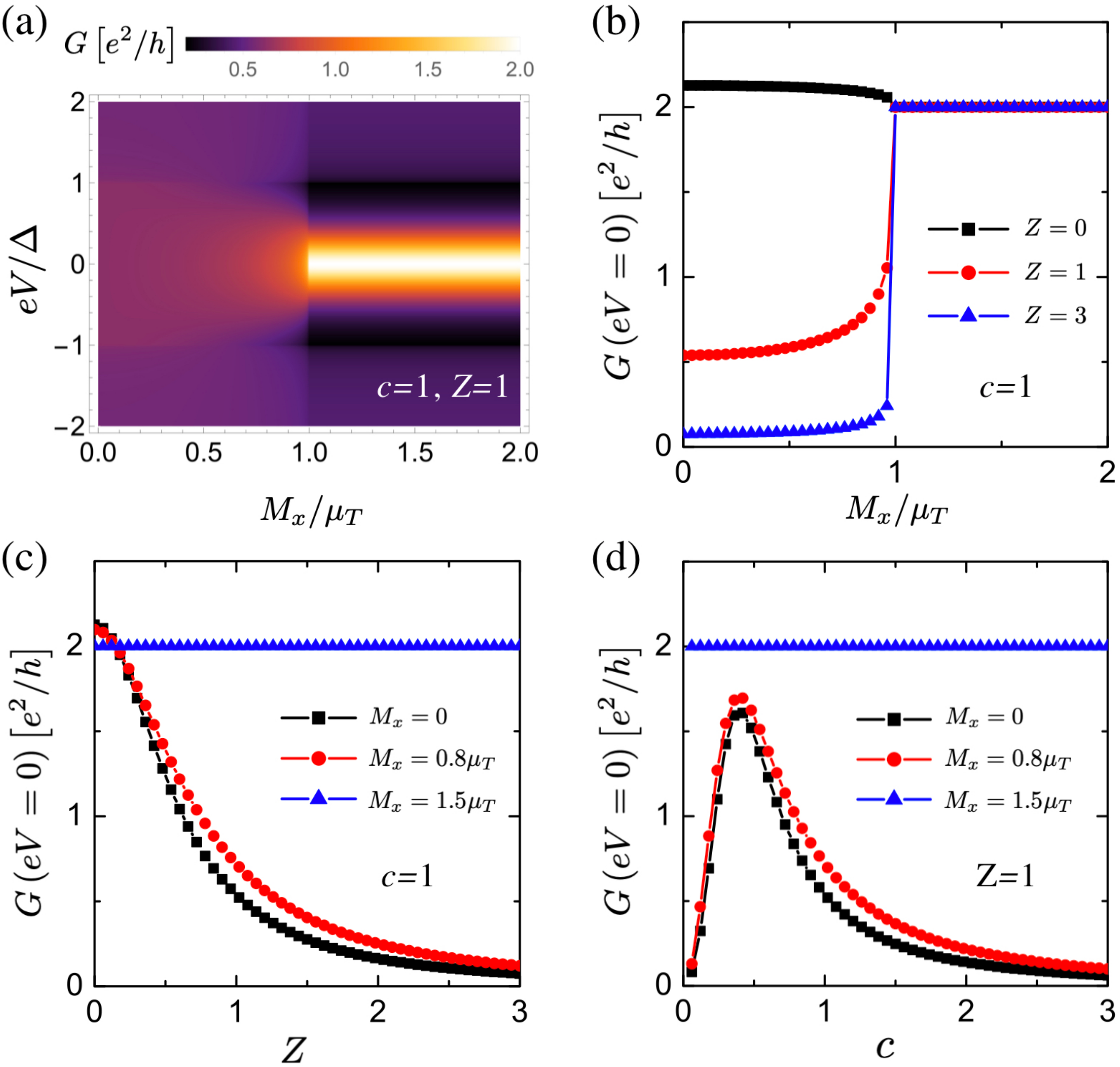}
	\caption{Zero-temperature conductance on the normal-metal terminal placed at the FS boundary. (a) The conductance spectra at zero temperature as a function of the energy $eV$ and magnetization $M_{x}$ for $c=1$ and $Z=1$. (b), (c), and (d) show the zero-bias conductance as a function of $M_{x}$, $Z$, and $c$, respectively. We set $c=1$ for (b) and (c), and $Z=1$ for (d). }
	\label{cond1}
\end{figure}

\section{III. FS junction}
We start with the FS junction [\cref{fig:01}(a)] and analyze the conductance probed by a normal-metal terminal positioned at the boundary between the magnetic and superconducting regions, where the \gls{mbs} is spatially localized. 
For $M_{x}<\mu _{T}$, the magnetic gap is buried beneath the Fermi level and there are propagating channels in the F region, whereas a magnetic gap starts to develop near the Fermi surface when $M_{x}>\mu_{T}$. The critical point between these two topological phases is thus $M_{x}=\mu _{T}$.  
This phase boundary clearly appears in the conductance dependence on the magnetization $M_{x}$, see \cref{cond1}(a). The conductance for $M_{x}>\mu _{T}$ clearly showcases the celebrated Majorana zero-bias peak with quantized height $2e^{2}/h$~\cite{Sengupta01,Bolech07,Akhmerov09,Tanaka09,Law09,Flensberg10,Ikegaya15,Ikegaya16,Bo16,Fleckenstein18,Fleckenstein18b}, and also features a dip at the gap edge. 
Reducing $M_{x}$ from $\mu _{T}$ towards $0$, the value of the \gls{zbp} loses its quantization and can either increase when $Z$ is small or decrease for large $Z$, see \cref{cond1}(b). 
\Cref{cond1}(c) and \cref{cond1}(d) show that the quantized value of the \gls{zbp} is immune to variations of the parameters $Z$ and $c$, respectively, when in the topological phase with $ M_{x}>\mu _{T}$. The robust $2e^{2}/h$ value indicates the presence of a single \gls{mbs}. 
However, in the trivial phase ($M_{x}<\mu _{T}$) the conductance dependence on $c$ and $Z$ is very different. 
First, for $Z=0$ the metallic probe is transparently coupled to the helical state and all the injected current at zero bias can either directly flow to the drain in F or undergo a perfect Andreev reflection on S. This results in a conductance value slightly bigger than $2e^2/h$, which can not be achieved in the absence of a tip when the bias is applied to the F region. For $Z\neq 0$, electrons incoming from the tip can backscatter, thus monotonically reducing the conductance as $Z$ increases. 
By contrast, the dependence on $c$ of the \gls{zbp} displays a non-monotonic behavior in the trivial phase. The conductance is suppressed for $c\rightarrow 0,\infty$, since the probe is decoupled from the edge state in these limits, and thus exhibits a maximum in between, whose position depends on both $c$ and $Z$~\cite{Soori20}. 

\begin{figure}
	\includegraphics[width=0.47\textwidth]{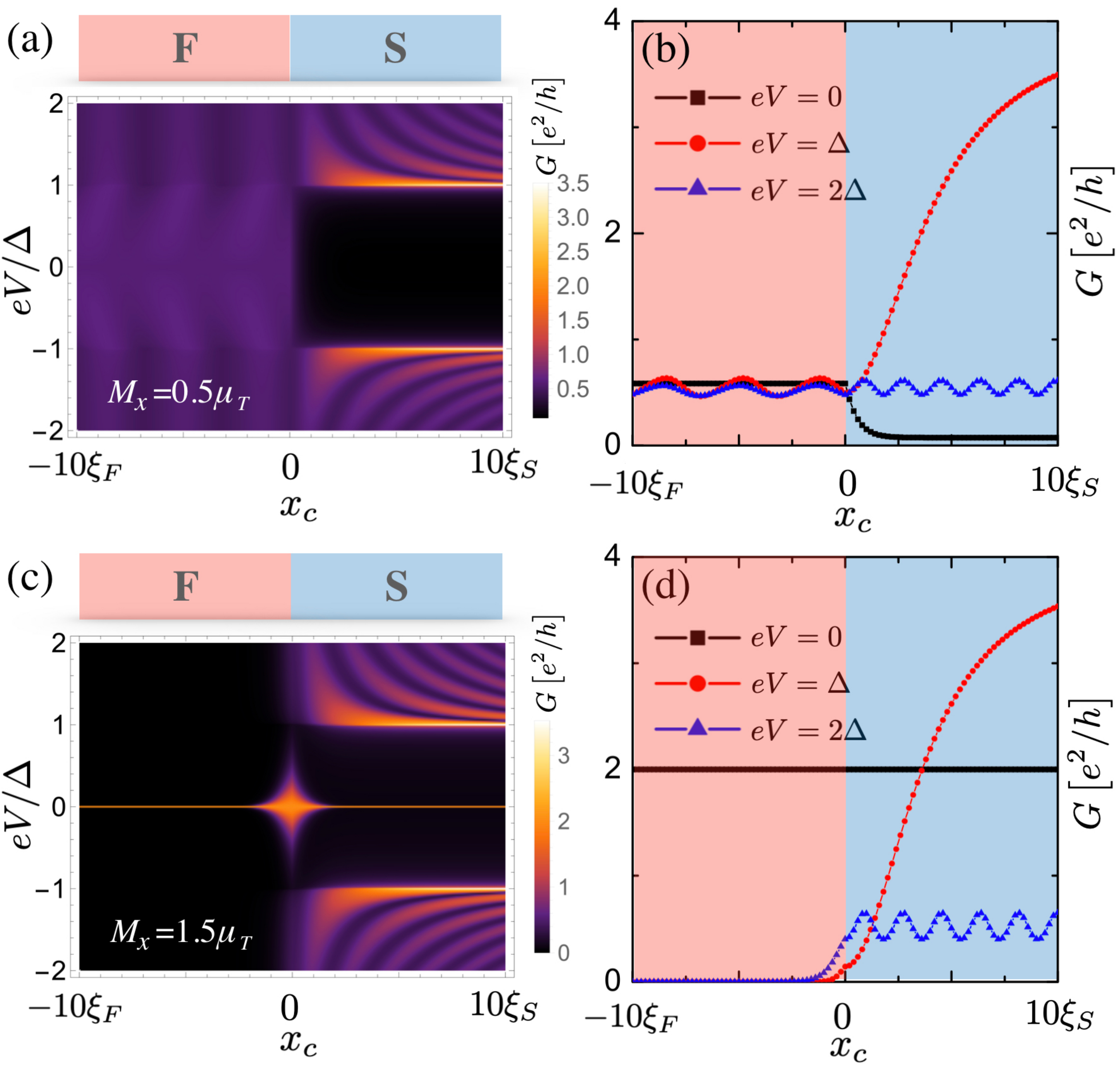}
	\caption{Spatial dependence of the probe conductance at zero temperature for the FS junction. 
		(a,b) In the trivial phase with $M_{x}/\mu _{T}=0.5$, (a) conductance spectra as a function of the bias $eV$ and the position $x_c$, and (b) three cuts corresponding to biases $eV=0$, $\Delta$ and $2\Delta$. 
		(c,d) Same as before for the topological phase with $M_{x}/\mu _{T}=1.5$. 
		The scale of the horizontal axis is different in each region since $\xi _{F}\ll \xi _{S}$. 
		In all cases, $c=1$ and $Z=1$. }
	\label{fig:space1}
\end{figure}

When the probe is at the FS boundary, we have shown a perfectly quantized \gls{zbp} in the topological phase $M_{x}>\mu_{T}$. However, this exact position may be challenging to reach in experiments, so we now consider the spatial dependence of the conductance as the probe moves along the \gls{qsh} edge. First, it is convenient to define two characteristic length scales: $\xi _{F}=\hbar v_{t}/\mu _{T}$ inside F and $\xi_{S}=\hbar v_{t}/\Delta $ in S. 
\Cref{fig:space1} shows the dependence on the coordinate $x$ of the conductance at the normal terminal probe. As expected, there is no quantized \gls{zbp} in the trivial phase ($M_{x}<\mu_{T}$), see \cref{fig:space1}(a) and \cref{fig:space1}(b). In this regime, the tip conductance in the F region features small oscillations around a constant value, while it displays a gapped profile with large peaks at $|eV|=\Delta$ in the S region, induced by the density of states in a uniform superconductor. 

By contrast, the topological phase ($M_{x}>\mu _{T}$) shown in %
\cref{fig:space1}(c) and \cref{fig:space1}(d) presents a remarkable result.
The zero bias conductance is not only quantized at the FS boundary, but also
maintains a constant value of $2e^{2}/h$ for \textit{every position} along
the \gls{qsh} edge. The emergence of such a robust \gls{zbp} seems exotic,
especially for $x_c\rightarrow +\infty $, i.e., several coherence lengths ($%
\xi _{S}$) away from the localized \gls{mbs}. We note that the \textit{bulk}
proximitized superconducting region ($x_c\rightarrow +\infty $) is fully
gapped at zero energy, without free quasiparticles and with the local
density of the \gls{mbs} greatly suppressed~\cite%
{Crepin14,Bo15,Burset15,Fleckenstein18,Keidel18}. 
To further analyze this counter-intuitive result, we examine the zero-bias scattering amplitudes for $x_c>0$ in the topological phase with $M_{x}>\mu_{T}$, namely, 
\begin{subequations}\label{eq:scat-amps_E0}
\begin{align}
	b_{\uparrow \uparrow }& =-\left[ c^{4}\left( i+2Z\right) ^{2}+\eta ^{2}%
	\right] \Xi ^{-1}, 
	\\
	a_{\uparrow \uparrow }& =-e^{-2ik_{T}x_c}\Omega _{+}\left[ c^{4}+\left(
2c^{2}Z-i\eta \right) ^{2}\right] \Xi ^{-1}, 
	\\
	b_{\uparrow \downarrow \left( \downarrow \uparrow \right) }& =\pm ie^{\pm
		2ik_{T}x_c}\Omega _{\mp }\left[ c^{2}\left( i+2Z\right) \pm i\eta \right]
	^{2}\Xi ^{-1}, 
	\\
	a_{\uparrow \downarrow \left( \downarrow \uparrow \right) }& =-i\left[
	c^{4}\left( 1+4Z^{2}\right) \mp 2c^{2}\eta +\eta ^{2}\right] \Xi ^{-1},
\end{align}
\end{subequations}
with $\Omega _{\pm }=M_{x}/( \pm i\mu _{T}+\sqrt{M_{x}^{2}-\mu _{T}}%
) $, $\Xi =2[ c^{4}\left( 1+4Z^{2}\right) +\eta ^{2}] $, $%
\eta =v_{t}/v_{N}$, $a_{\downarrow \downarrow }=-a_{\uparrow \uparrow
}^{\ast }$ and $b_{\downarrow \downarrow }=b_{\uparrow \uparrow }$. 
The spacial dependence at zero energy appears as global phase factors $e^{\pm 2ik_{T}x_c}$, so the resulting conductance is thus quantized and independent of the position $x_c$. 
This effect provides an interesting signature of the existence of \gls{mbs} in a QSHI. 
\begin{figure}
	\centering{}
	\includegraphics[width=0.47\textwidth]{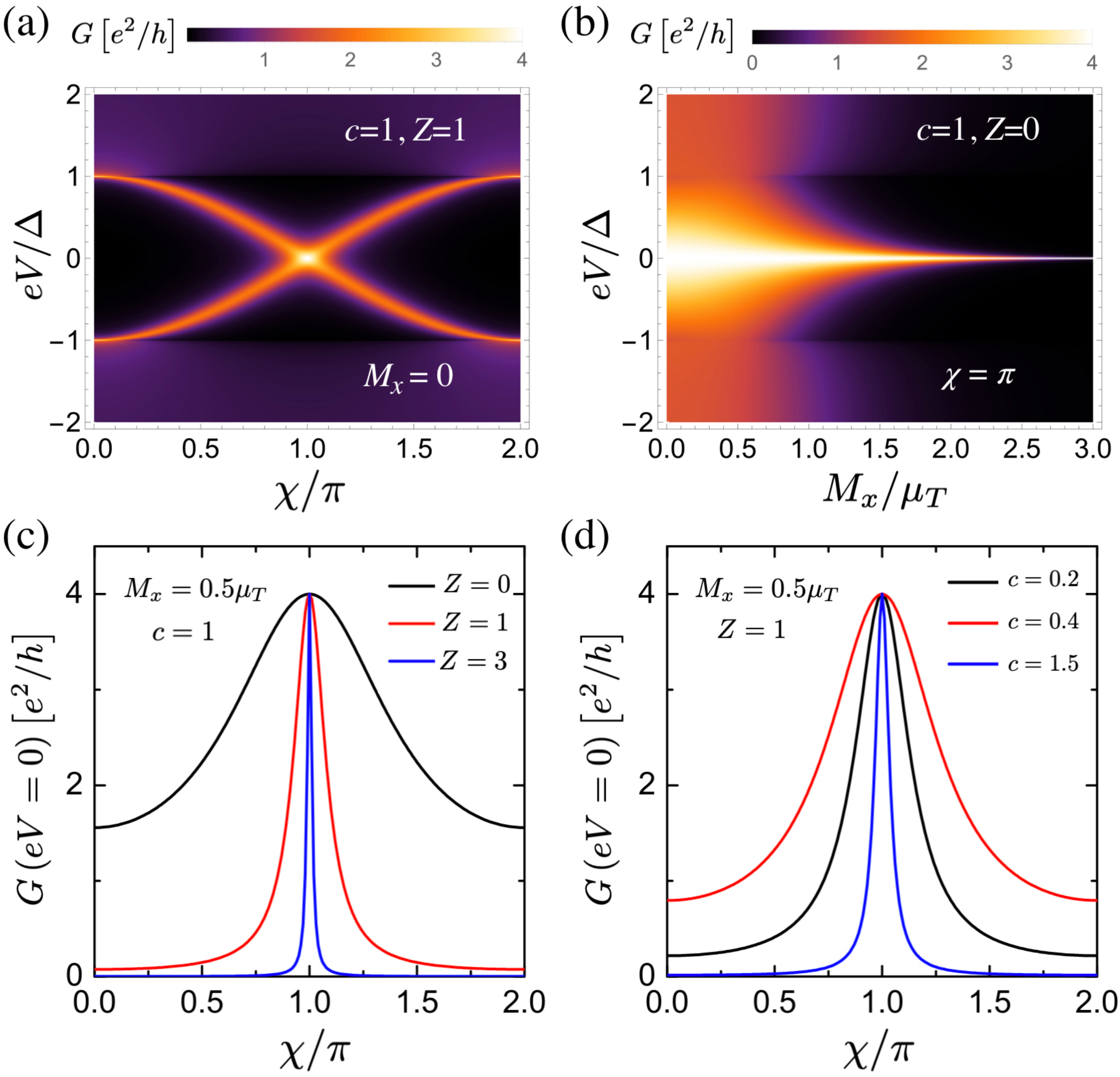}
	\caption{Zero-temperature conductance of a normal-metal probe placed in the middle of an SFS junction. 
		(a) Conductance versus applied voltage and phase difference $\chi$ for $M_x=0$ and $Z=c=1$. (b) For fixed $\chi=\pi$, conductance versus $eV$ and $M_x$, with $Z=0$ and $c=1$. 
		(c,d) Zero bias conductance as a function of $\chi$ for $M_x=0.5\mu_T$ and several values of $Z$ (c) or $c$ (d). We set $c=1$ for (c) and $Z=1$ for (d). }
	\label{fig:03}
\end{figure}

\section{IV. SFS junction}
We now connect the normal-metal terminal probe to the SFS junction. First, we place it in the middle of F and study the conductance as a function of  $M_{x}$ and the superconducting phase difference $\chi$ in \cref{fig:03}. At zero magnetization [\cref{fig:03}(a)], a resonance peak inside the superconducting gap indicates the presence of a pair of \glspl{mbs} exhibiting a protected crossing at $\chi=\pi$. 
The robustness of the crossing point is shown in \cref{fig:03}(b), where the zero biased conductance is quantized with $G=4e^{2}/h$ for $\chi=\pi$ and arbitrary $M_x$, similar to works in other topological Josephson junctions~\cite{Stefano14,Wang15,Riwar17}. Indeed, the \gls{qsh}-mediated Josephson junction is in the topological phase independently of the magnetization~\cite{Fu09,Tanaka09}. 
For nonzero $eV$, the transition between a dominant superconducting gap into a dominant magnetic one can be seen around $M_x=\mu_T$; as the magnetic gap becomes dominant for $M_x>\mu_T$, the conductance is suppressed. 
\Cref{fig:03}(c) and \cref{fig:03}(d) show the robustness of the \gls{zbp} at $\chi=\pi$ against the coupling parameters $c$ and $Z$. 

\begin{figure}
	\includegraphics[width=0.47\textwidth]{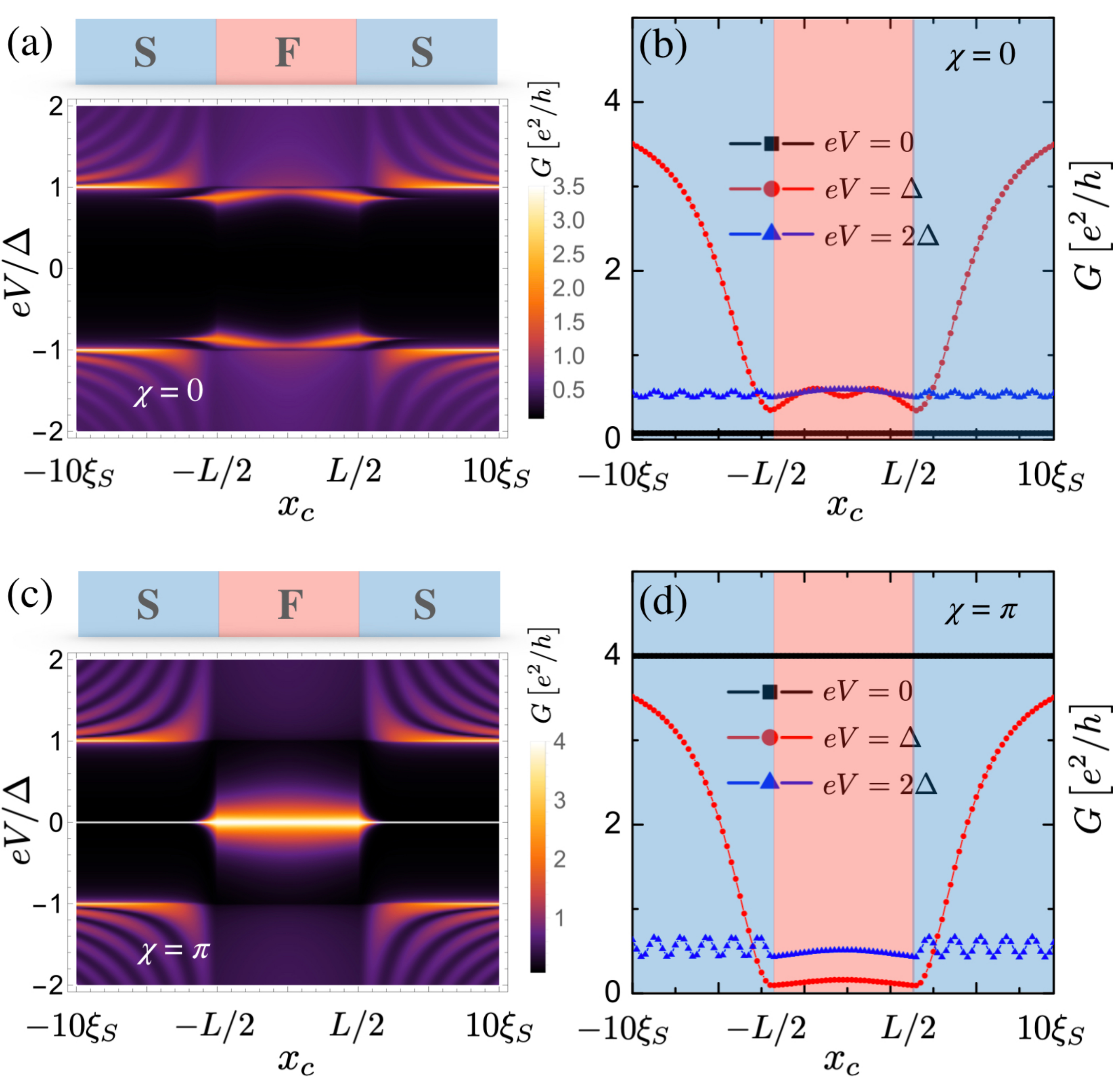}
	\caption{Spatial dependence of the probe conductance at zero temperature for an SFS junction. 
		(a, b) For $\chi=0$, (a) conductance spectra as a function of bias and position, and (b) three cuts corresponding to biases $eV=0$, $\Delta$ and $2\Delta$. 
		(c,d) Same as before for $\chi=\pi$. In all cases, we set $M_{x}=0.5\mu _{T}$, $Z=1$, and $c=1$. }
	\label{fig:space2}
\end{figure}
Next, we compare the spacial dependence of the probe conductance for $\chi=0$ [\cref{fig:space2}(a) and \cref{fig:space2}(b)] and $\chi=\pi$ [\cref{fig:space2}(c) and \cref{fig:space2}(d)]. For $\chi=0$, there is no \gls{zbp} in the SFS junction since the \glspl{mbs} have merged with the continuum. 
By contrast, at the crossing $\chi=\pi$, the zero-bias conductance becomes perfectly quantized to $4e^{2}/h$. As it was the case for the FS junction in the topological phase, the normal probe conductance remains quantized for any position across the whole SFS junction. However, the quantized \gls{zbp} is now present for any value of $M_{x}$, and is broadened in the F region, cf. \cref{fig:03}(c). Analytically, when the probe is on the left S side ($x_{c}<-L/2$), the coefficients are obtained as
\begin{subequations}
	\label{eq:scat-amps_SNS)}
	\begin{align}
		b_{\uparrow \uparrow (\downarrow \downarrow )}& =-\left( 1+e^{i\chi }\right)
		\Upsilon _{1}/\left( \Upsilon _{2}\pm \Upsilon _{3}\right) , \\
		a_{\downarrow \uparrow \left( \uparrow \downarrow \right) }& =\mp \Upsilon
		_{4}/\left( \Upsilon _{2}\pm \Upsilon _{3}\right) ,
	\end{align}
\end{subequations}
and $b_{\uparrow \downarrow }=b_{\downarrow \uparrow }=a_{\uparrow \uparrow
}=a_{\downarrow \downarrow }=0$, with $\Upsilon _{1}=c^{4}\left(
i+2Z\right) ^{2}+\eta ^{2}$, $\Upsilon _{2}=\left( 1+e^{i\chi }\right)
\left( c^{4}+4c^{4}Z^{2}+\eta ^{2}\right) $, $\Upsilon _{3}=2c^{2}\left(
e^{i\chi }-1\right) \eta e^{\Delta \left( L+2x_{c}\right) /(\hbar v_{t})}$
and $\Upsilon _{4}=2ic^{2}\left( 1+e^{i\chi }\right) \eta -i\Upsilon _{3}$.
As the phase bias is $\chi =\pi $, $b_{\sigma \sigma }$ becomes $0$ while $%
a_{\downarrow \uparrow }=a_{\uparrow \downarrow }=i$ exhibiting perfect
Andreev reflection, regardless of the position of probe. 

\section{V. Temperature effect}
\begin{figure}
	\includegraphics[width=0.47\textwidth]{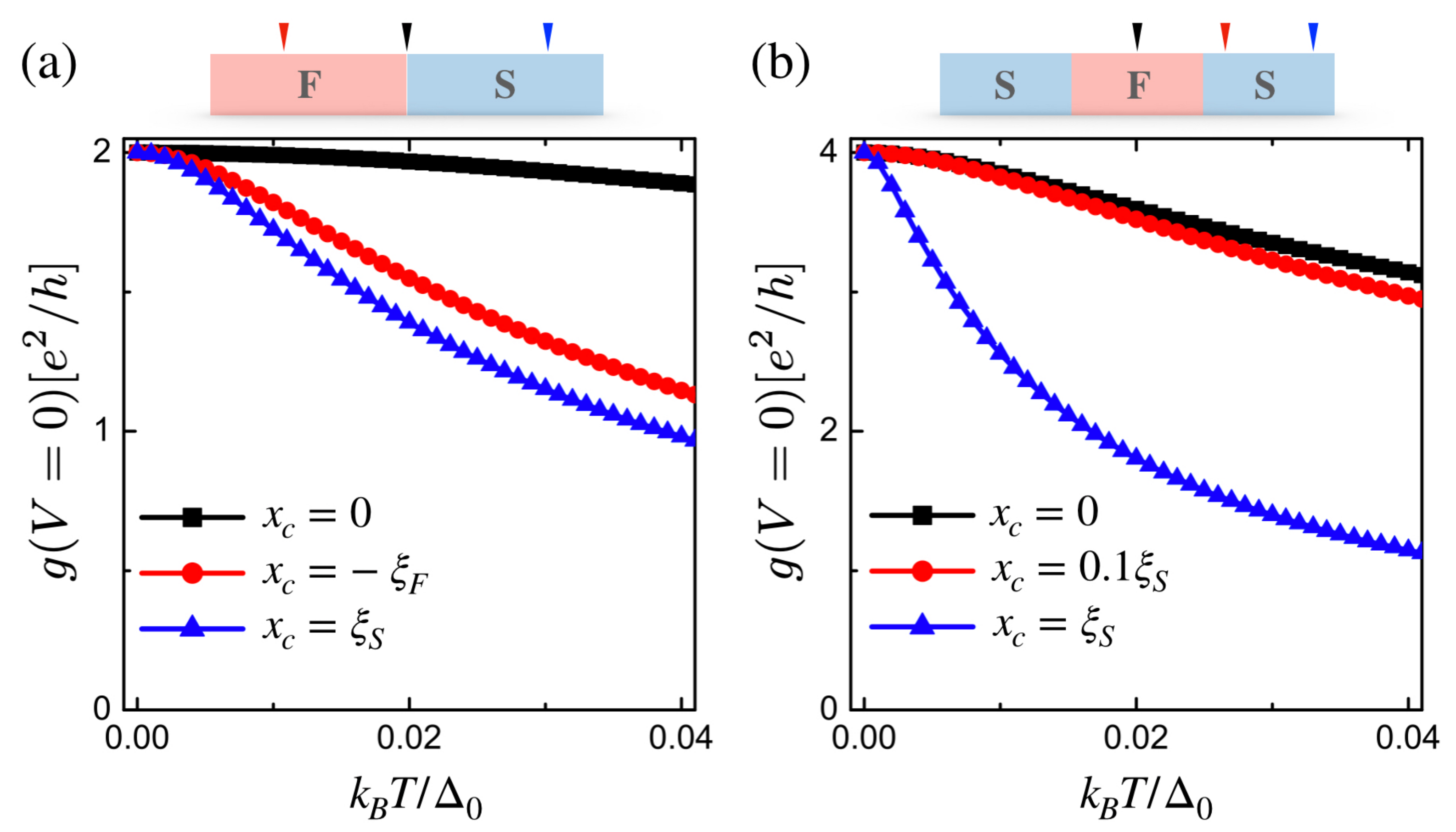}
	\caption{Zero-biased conductance as a function of temperature $T$ for different positions of the probe. (a) The probe is placed at $x_c=0, -\xi_F$, and $\xi_S$ in the FS junction. The parameters of the FS junction are the same as Fig. \ref{fig:space1}(c). (b) The probe is placed at $x_c=0, 0.1\xi_S$, and $\xi_S$ in the SFS junction with the same parameters as in Fig. \ref{fig:space2}(c). $\Delta_0$ is the superconducting gap at zero temperature. } 
	\label{fig:Temp}
\end{figure}

So far we have reported that the ZBCP always sticks to the quantized value
at zero temperature. To obtain a result that can be directly compared to
experiments, we next remark on the temperature effect. At zero temperature,
we find that the topological ZBCP becomes sharper as the probe is placed far
from the localized MBSs, see \cref{fig:space1}(c) and \cref{fig:space2}(c).
This indicates that the height of the ZBCP would be lowered at finite
temperatures due to the thermal broadening. We calculate the conductance at
finite temperature $T$ as
\begin{equation}
	g(V) =\int_{-\infty }^{+\infty }\mathrm{d}\varepsilon G\left(
	\varepsilon \right) \left[ 4k_{B}T\cosh ^{2}\left( \frac{\varepsilon -eV}{%
		2k_{B}T}\right) \right] ^{-1},
\end{equation}%
where $G\left( \varepsilon \right) $ is the conductance at zero temperature
given in \cref{eq13}. We find that the ZBCP height does not change
significantly by increasing the temperature $T$ as the probe tip is placed
at $x_{c}=0$, i.e., the interface of the FS junction or the middle of the
SFS junction, as shown in \cref{fig:Temp}. However, a finite temperature has
a drastic suppression effect on the ZBCP height as the tip moves far away
from $x_{c}=0$. 
This is consistent with recent experiments where a robustZBCP was found around the FS interface at finite temperature~\cite{Jack_2019}.  
Our results show that the weakened ZBCP is still observable if the probe
is within the localization length of the MBSs at low temperature, e.g., $%
x_{c}=0.1\xi _{S}$ (red line in \cref{fig:Temp}(b)) in the SFS junction.
Consequently, the unusual spatial independence of the ZBCP found in our
calculation could be experimentally observed in low temperature measurements.

\section{VI. CONCLUSION}
We studied the conductance afforded by a normal-metal probe which directly contacts the helical edge modes of a quantum spin Hall insulator. 
We found a robust quantized zero-biased conductance peak in both FS and SFS junctions indicating the presence of Majorana bound states at each FS boundary. The conductance quantization is robust under variations of the parameters controlling the coupling between the tip and the helical edge state. 
Moreover, we found that the zero bias conductance quantization remains unchanged as we moved the probe along the edge, remarkably, even for distances much larger than the localization length of the Majorana states. 
This result can not be simply explained by the local density of states in the tunneling model~\cite{Bours18,Bours19,Blasi20}. 
Our analytical results suggested that the spatial independence of the zero-bias tip conductance results from a coherent coupling to the zero-energy MBSs.  Finally, we discuss the temperature effect on the ZBCP to estimate the quality of the conductance quantization in actual experiments.
Our proposal for the observation of Majorana states using a metallic probe is within reach of recent experimental advances implementing hybrid superconductor and magnetic structures on the quantum spin Hall insulator~\cite{Knez12,Hart14,Pribiag15,Bocquillon17, Deacon17,Sajadi_2018,Fatemi_2018,Wu18,Jack_2019}.

\section{Acknowledgments}
B. L. acknowledges support from the National Natural Science Foundation of China (project 11904257) and the Natural Science Foundation of Tianjin (project 20JCQNJC01310).
P. B. acknowledges support from the Spanish CM ``Talento Program'' Project no.~2019-T1/IND-14088 and the Spanish Ministerio de Ciencia e Innovaci\'on Grant no.~PID2020-117992GA-I00. Y. T. was supported by Grant-in-Aid for Scientific Research on Innovative Areas and Grant-in-Aid for Scientific Research B (Grant No. JP18H01176) from the Ministry of Education, Culture, Sports, Science, and Technology, Japan (MEXT). Y. T. was also supported by Grant-in-Aid for Scientific Research A (KAKENHI Grant No. JP20H00131) and the JSPS Core-to-Core program Oxide Superspin International Network.

\bibliography{SpinSC}

\end{document}